\documentclass[preprint,showpacs,superscriptaddress,aps]{revtex4-1}
\usepackage{amssymb}
\usepackage{amsmath,amssymb,graphicx}
\usepackage{graphicx}
\usepackage{dcolumn}
\usepackage{bm}
\def\be{\begin{equation}}
\def\ee{\end{equation}}
\def\bea{\begin{eqnarray}}
\def\eea{\end{eqnarray}}

\newcommand{\beq}{\begin{equation}}
\newcommand{\eeq}{\end{equation}}
\newcommand{\beqa}{\begin{eqnarray}}
\newcommand{\eeqa}{\end{eqnarray}}

\def\lsim{\raise0.3ex\hbox{$\;<$\kern-0.75em\raise-1.1ex\hbox{$\sim\;$}}}
\def\gsim{\raise0.3ex\hbox{$\;>$\kern-0.75em\raise-1.1ex\hbox{$\sim\;$}}}

\usepackage{pstricks}

\begin{document}

\title{New Physics signature in $D^0 (\bar{D}^0)\to f$  effective width asymmetries}

\author{David Delepine}
\email{delepine@fisica.ugto.mx}
\affiliation{{\fontsize{10}{10}\selectfont{Division de Ciencias e
Ingenier\'ias,  Universidad de Guanajuato, C.P. 37150, Le\'on,
Guanajuato, M\'exico.}}}

\author{Gaber Faisel}
\email{gaberfaisel@sdu.edu.tr}

\affiliation{{\fontsize{10}{10}\selectfont{Department of Physics,
Faculty of Arts and Sciences, S\"uleyman Demirel University,
Isparta, Turkey 32260.}}}

\author{ Carlos A. Ramirez}
\email{jpjdramirez@yahoo.com}
\affiliation{{\fontsize{10}{10}\selectfont{Depto. de F\'isica,
Universidad de los Andes, A. A. 4976-12340, Bogot\'a, Colombia.}}}

\begin{center}
\date{\today}
\begin{abstract}
Violation of charge conjugation-parity ($\rm CP$) symmetry plays a
major rule in the dominance of matter in our universe.  A kind of
$\rm CP$ violation results from the asymmetry of the life time
measured in $M^0$ and $\bar M^0$, here $M$ is a heavy meson,
decays to final states which is referred in the literature  as
$A_{\Gamma}^f$. In this paper, we give an estimation of the upper
bound on  $|A_{\Gamma}^f|$ for the Cabibbo Favored $D^0
\rightarrow K^- \pi^+$ decay process in different models. We show
that in the standard model, $|A_{\Gamma}^f| \lesssim\mathcal{O}
(10^{-10})$. Recently a bound on $A_{\Gamma}^f$ has been obtained:
$(A^f_{\Gamma})^{Exp.}= (1.6 \pm 1)\times 10^{-4}$. This result
motivates further studies on $A_{\Gamma}^f$ in beyond standard
model physics. In the framework of two Higgs doublet model with
generic Yukawa structure, we show that $|A_\Gamma^{f}|\lesssim
\mathcal{O} (10^{-7})$ which is several orders of magnitude
smaller than the current experimental value. Finally, in the
framework of left-right symmetric models in which the mixing
between the left and the right gauge bosons is allowed and the
left-right symmetry is not manifest at unification scale, we find
that $A_{\Gamma}^f$ can be as large as
$|A_{\Gamma}^f|\lesssim\mathcal{O} (10^{-5})$ which is one order
of magnitude smaller than the experimentally measured value by
LHCb collaborators.
\end{abstract}

\end{center}

\maketitle
\flushbottom

\section{Introduction}

Symmetries play an important rule in particle physics. In the
standard model (SM), masses of quarks, charged leptons and weak
gauge bosons can be attributed to the breaking of the electroweak
symmetry. On the other hand, the difference between the decay
rates of particle and its antiparticle can be an indication of
direct violation of charge-parity (CPV) symmetry. Weak decays of
hadrons can serve as a probe for CPV. This remark can be
understood as in SM CPV originates from the presence of complex
couplings in the Cabibbo-Kobayashi-Maskawa (CKM) matrix which
appears only in the quark sector in the interactions of quarks and
the charged weak gauge bosons $W^\pm$
\cite{Cabibbo:1963yz,Kobayashi:1973fv}.

Direct CPV has been confirmed in the weak decays of K and B mesons
\cite{Christenson:1964fg,Aubert:2004qm,Aaij:2013iua,Aaij:2012kz}.
On the other hand, the remarkable experimental progress in D
mesons has leaded to the observation of $D^0-D^0 $ meson mixing
\cite{Aubert:2007wf,Aaltonen:2007ac,Staric:2007dt,Aaij:2012nva}
and measurements of direct CP asymmetries in D mesons decays, with
precision of O($10^{-3})$ \cite{Aaij:2016dfb}. A sensitive probe
of CP violation in the weak decays of $D^0$ meson is given by the
direct CP asymmetry difference, $\Delta A_{CP}$, between $D^0\to
\pi^+\pi^-$ and $D^0\to K^+K^-$ which can be expressed as \be
\Delta A_{CP} =  A_{CP}(K^-K^+) - A_{CP}(\pi^-\pi^+) \ee The first
observation of  $\Delta A_{CP}$ was reported in 2011 by the LHCb
Collaboration \cite{Aaij:2011in} and later confirmed by CDF
\cite{Collaboration:2012qw}  and Belle \cite{Ko:2012px}. Recently,
the LHCb collaboration has presented new measurements at Moriond
2019 and the combined value  with previous LHCb results leads to
\cite{Aaij:2019kcg} \be \Delta A_{CP} = (-15.6 \pm 2.9) \times
10^{-4} \ee which is $5.3$ standard deviations away from zero and
thus confirm direct CPV in these particular weak decays of $D^0$
mesons. This progress motivates further search and further studies
of CP violation in D meson decays.

Indirect CP violation has been searched also in the decays of
$D^0$ and $\bar D^0$ to final states $K^+K^-$, $\pi^+\pi^-$ and
$K^-\pi^+$ modes \cite{Aaij:2017idz}. This kind of CP violation
results from the asymmetry of the life time measured in $D^0$ and
$\bar D^0$ decays to the same final states or equivalently the
asymmetry in effective decay widths and usually denoted by
$A_{\Gamma}$. The latest measurements are given as
\cite{Aaij:2017idz}
\begin{eqnarray}
A_{\Gamma}(K^-K^+)&=& \left( -0.30\pm 0.32\pm 0.10 \right) \times 10^{-3}  \nonumber\\
A_{\Gamma}(\pi^-\pi^+)&=& \left(0.46 \pm 0.58 \pm 0.12 \right) \times 10^{-3}\nonumber\\
A_{\Gamma}(K^-\pi^+)&=& (0.16 \pm 0.10)\times 10^{-3}
\end{eqnarray}
The first two processes are single Cabibbo suppressed  (SCS) and
the Standard Model (SM) contributions to these asymmetries are
expected to be of order
$10^{-4}$\cite{Bianco:2003vb,Bobrowski:2010xg,1,2}. On the other
hand, the last one is Cabibbo favored (CF) with very suppressed
direct CP asymmetry in the framework of the SM
\cite{Delepine:2012xw}. So any observation of CPV in these CF
channels will be a strong hint for New Physics. At present time,
all results are compatible with no direct  or indirect
CPV\cite{Aaij:2014afa,Aubert:2008yd,LHCb:2013dka,
Aaij:2014gsa,Aaij:2013wda,Aaij:2015yda,Aaij:2017idz}.

In the literature, the study of the interference between direct
CPV and mixing has been performed through the introduction of a
non-universal weak phase defined as $\delta_f \equiv -\arg(\bar
A_f/ A_f)$ where $A_f$ is the $A(D^0 \to f)$ amplitude
\cite{Kagan:2009gb,Grossman:2012eb,Dighe:2013epa}. It is important
to notice that this weak phase is irrelevant for the direct CPV as
direct CPV is proportional to $|A_f|^2-|\bar A_f|^2$. To get non
vanishing direct CPV it is necessary to write the amplitude,
$A_f$, as a sum of at least two amplitudes with different relative
weak and strong phases.

The mass eigenstates of the neutral $D$ mesones, denoted as
$|D_{1,2}>$ with masses (total widths) $m_{1,2}$ ($\Gamma_{1,2}$),
are linear combinations of the flavor eigenstates $|D^0>$ and
$|\overline{D}^0>$ and can be defined as follows:
\begin{equation}
|D_{1,2}>= p|D0> \pm q|\overline{D}^0>
\end{equation}
with the imposed normalization condition  $|p|^2+|q|^2=1$. We
consider the decay modes $D^0\to f(\bar f)$ and $\overline{D}^0\to
f(\bar f)$, with $f \equiv K^-\pi^+$ and $\bar f \equiv K^+\pi^-$.
These modes are examples of $D^0$ and $\bar D^0$ decays to final
non-CP eigenstate modes. In the following we denote the decay
amplitudes as, $A_f=A(D^0\to f)$, $\bar A_f=A(\overline D^0\to
f)$, $A_{\bar f} =A(D^0\to \bar f)$ and $\bar A_{\bar
f}=A(\overline D^0\to \bar f)$.  Moreover, we follow
Ref.\cite{Grossman:2006jg} and express the amplitudes as
\beqa\label{fouramp} A_f&=&A^T_{f}
e^{+i\phi^T_{f}}[1+r_fe^{i(\delta_f+\phi_f)}],\qquad
\overline{A}_{\overline f}= A^T_{f}
e^{-i\phi^T_{f}}[1+r_fe^{i(\delta_f-\phi_f)}],\nonumber\\
A_{\overline f}&=& A^T_{\overline f}
e^{i(\Delta_{f}+\phi^T_{\overline f})} [1+r_{\overline f}
e^{i(\delta_{\overline f}+\phi_{\overline f})}] ,\qquad
\overline{A}_f=A^T_{\overline f} e^{i(\Delta_{f}-\phi^T_{\overline
f})} [1+r_{\overline f }e^{i(\delta_{\overline f}-\phi_{\overline
f })}]. \eeqa here $A^T_{f} e^{ i\phi^T_{f}}$ and $A^T_{\overline
f} e^{i(\Delta_{f}+\phi^T_{\overline f})}$ represent SM tree-level
contribution, the phases $\phi_f^T$ and $\phi^T_{\overline f}$
represent weak CP violating phases while $\Delta_f$ represent
strong CP conserving one all generated at tree-level. It should be
noted that, upon neglecting small terms of order
$|(V_{ub}V_{cb})/(V_{us}V_{cs})|\sim10^{-3}$,
$\phi^T_f=\phi^T_{\overline f}$.  In Eq.(\ref{fouramp}), the
quantities $r_f$ and $r_{\bar f}$ express the relative magnitudes
of subleading contributions that can arise from new physics or
from SM amplitudes with suppressed CKM factors. Moreover, the
phases $\phi_{f,\overline f}$ and $\delta_{f,\overline f}$ denote
the relative weak and strong CP violating phases respectively that
account for the difference between the phases generated by the
subleading contributions and the tree-levels ones.

Using the general formalism for $D^0-\overline{D}^0$ mixing, it is
possible to compute the widths as a function of time. The
time-dependent decay rates can be expressed as
\cite{Nir:2005js,Grossman:2006jg}
\begin{eqnarray}
\Gamma(D^0(t)\to f) &=& \big|A_f(t)\big|^2 =  \big| g_+(t)
A_f(t)+{q\over p}g_-(t)\bar A_f(t)\big|^2  = \big|A_f(t)\big|^2
\big[ |g_+(t)|^2+|g_-(t)\lambda_f|^2 \nonumber\\&+&2{\rm
Re}\big(g_+^*(t) g_-(t)\lambda_f\big)\big] = {e^{-\Gamma t}\over
2}\big|A_f(t)\big|^2 \big[ (1+|\lambda_f|^2)\cosh(y\Gamma
t)+(1-|\lambda_f|^2)\cos (x\Gamma t)\nonumber\\&+&2{\rm Re}
\lambda_f
\sinh (y\Gamma t) -2{\rm Im} \lambda_f \sin (x\Gamma t) \big],\nonumber \\
\Gamma(\overline{D}^0 \rightarrow f) (t) &=& \big|\bar
A_f(t)\big|^2= \big| {p\over q}g_-(t)A_f(t) +g_+(t)\bar A_f(t)
\big|^2  = \big|\bar A_f(t)\big|^2 \big|g_+(t)+
g_-(t)\lambda^{-1}_f\big|^2
\nonumber \\
&=&   \big|\bar A_f(t)\big|^2 \big[
|g_+(t)|^2+|g_-(t)\lambda^{-1}_f|^2+2{\rm Re}\big(g_+^*(t)
g_-(t)\lambda^{-1}_f\big) \big]
\nonumber \\
&=&{e^{-\Gamma t}\over 2} \big|\bar A_f(t)\big|^2 \big[
\big(1+|\lambda^{-1}_f|^2\big)\cosh (y\Gamma
t)+\big(1-|\lambda^{-1}_f|^2\big)\cos (x\Gamma t)+2{\rm
Re}\big(\lambda^{-1}_f\big) \sinh (y\Gamma t) \nonumber \\
&-& 2{\rm Im}\big(\lambda^{-1}_f\big) \sin (x\Gamma t)
\big].\label{timee}
\end{eqnarray}
here $\Gamma = (\Gamma_1+\Gamma_2)/2$ is the mean $D^0$ width and
\begin{eqnarray}
g_{\pm}(t) &=& {1\over 2}\bigg({\rm e}^{-i m_2
t-\frac{1}{2}\Gamma_2 t}\pm {\rm e}^{-i m_1 t-\frac{1}{2}\Gamma_1
t} \bigg)\nonumber\\
x&=&\frac{\Delta m}{\Gamma}, \  y=\frac{\Delta{\Gamma}}{2\Gamma}\nonumber\\
\lambda_f&=&{q\bar A_f\over p A_f},
\end{eqnarray}
The experimental values for the mixing and CPV parameters in $D$
neutral mesons are given as \cite{Amhis:2014hma}
\begin{eqnarray}
x&=&0.41_{-0.15}^{+0.14} \% \\
y&=&0.63_{-0.08}^{+0.07} \% \\
a_{CP}^D&=& -0.71_{-0.95}^{+0.92} \\
|q/p|&=&0.93_{-0.08}^{+0.09} \\
\phi &=& -8.7_{-9.1}^{+8.7} \label{exp}\end{eqnarray} where the
fit assuming all floating parameters is used and $\phi=arg(q/p)$
is expressed in degree.

The expressions of the time-dependent decay rates into a final
state $\bar f$ can be obtained via the substitutions $f\to \bar f$
in the above expressions \cite{Grossman:2006jg}. Due to the small
values of the mixing parameters x and y and $|\lambda_f| \ll 1$
and $|\lambda^{-1}_{ \bar f}|\ll 1$, these approximations are
experimentally confirmed for the decay modes under consideration
\cite{Bergmann:2000id}, one can expand the expressions of the
time-dependent decay rates of $D^0\to f$ and $\overline{D}^0\to
\bar f$ and  keep the terms up to first order in time. Thus, we
get
\begin{eqnarray}
\Gamma\big[D^0(t)\to f\big] & \simeq & {\rm e}^{-\Gamma
t}\left|A_f\right|^2 \left\{1+ \bigg[y\, Re(\lambda_{f}) - x \, Im
(\lambda_{f})\bigg]\Gamma t \right\} \simeq \left|A_f\right|^2
{\rm e}^{-t\,\hat\Gamma_{D^0\to f}}
\nonumber \\
\Gamma \big[\overline D^0(t)\to {\overline f}\big] &\simeq & {\rm
e}^{-\Gamma t}\left|\bar A_{\overline f}\right|^2 \left\{1+
\bigg[y\, Re(\lambda^{-1}_{\overline f}) - x \, Im
(\lambda^{-1}_{\overline f})\bigg]\Gamma t \right\}\simeq
\left|\bar A_{\overline f}\right|^2 {\rm e}^{
-t\,\hat\Gamma_{\overline{D}^0\to \overline f}}
\label{wid111}\end{eqnarray}

Upon defining \cite{Grossman:2006jg,Kagan:2009gb}
\beq\lambda_f\equiv\frac qp\frac{\overline{A}_f}{A_f} =-R_m R_f
e^{i(\Phi+\Delta_f+\delta \phi_{\lambda_f})},\qquad
\lambda_{\overline f}\equiv\frac qp\frac{\overline{A}_{\overline
f}}{A_{\overline f}} =-R_mR_f^{-1}e^{i(\Phi-\Delta_f+\delta
\phi_{\lambda_{\overline f}})}.\label{lfs1} \eeq where
$R_m\equiv|q/p|$, $\Phi =\phi - \phi^T_f-\phi^T_{\overline f}$,
$R_f\equiv |\frac{\overline{A}_f}{A_f}|$. The phases
$\delta\phi_{\lambda_f}$ and $\delta\phi_{\lambda_{\overline f}}$,
to first order in $r_f$ and $r_{\overline f}$, are given as
\cite{Kagan:2009gb} \bea \delta\phi_{\lambda_f} &=& - r_f
\sin(\delta_f+\phi_f)+r_{\overline f} \sin(\delta_{\overline
f}-\phi_{\overline
f})\nonumber\\
\delta\phi_{\lambda_{\overline f}} &=& -r_{\overline f}
\sin(\delta_{\overline f}+\phi_{\overline f})+r_f
\sin(\delta_f-\phi_f) \eea Using the above definitions of
$\lambda_f$ and $\lambda_{\overline f}$, we find that the
effective widths $\hat\Gamma_{D^0\to f}$ and
$\hat\Gamma_{\overline{D}^0\to \overline f}$ in Eq.(\ref{wid111})
can be expressed as
\begin{eqnarray}
\hat\Gamma_{D^0\to f} &=&  \Gamma \left[1+R_m R_f\left( y\cos
\phi_{\lambda_f}-x\sin \phi_{\lambda_f}\right) \right]
\nonumber \\
\hat\Gamma_{\overline{D}^0\to \overline f} &=& \Gamma
\left[1+\frac{R_f}{R_m}\left(y\cos\phi_{\lambda_{\overline
f}}+x\sin\phi_{\lambda_{\overline f}}\right)\right]\label{efw}
\end{eqnarray}
here $\phi_{\lambda_f}$ and $\phi_{\lambda_{\overline f}}$ are the
arguments of $-\lambda_f$ and $-\lambda_{\overline f}$
respectively. Now, one can define the following CP observable, the
asymmetry in effective decay widths $A_{\Gamma}^f$, for the $D^0$
and $\overline D^0$ decays to final two-body non-CP eigenstate
mode $f$ \cite{Grossman:2006jg}:
\begin{eqnarray}
A_\Gamma^f &=& \frac{\hat\Gamma_{\overline{D}^0\to \overline
f}-\hat\Gamma_{D^0\to f}} {2\,\Gamma}
\end{eqnarray}
Thus, using Eq.(\ref{efw}), we obtain
\begin{eqnarray}
A_\Gamma^f &=&  {1\over 2}
\frac{R_f}{R_m}\left(y\cos\phi_{\lambda_{\overline
f}}+x\sin\phi_{\lambda_{\overline f}}\right) -{1\over 2} R_m
R_f\left( y\cos \phi_{\lambda_f}-x\sin \phi_{\lambda_f}\right)
\label{efw3}
\end{eqnarray}
Upon substitution of the expressions of $\phi_{\lambda_f}$ and
$\phi_{\lambda_{\overline f}}$  we get
\begin{eqnarray}
 A_\Gamma^f &=& \frac{R_f}{2} \bigg\{\bigg[R_m \bigg(x \cos
(\Delta_f+\delta \phi_{\lambda_f}) + y \sin(\Delta_f+\delta
\phi_{\lambda_f}) \bigg)+R^{-1}_m \bigg(x \cos (\Delta_f-\delta
\phi_{\lambda_{\overline f}}) + y \sin(\Delta_f-\delta
\phi_{\lambda_{\overline f}}) \bigg)\bigg] \nonumber\\&\times&
\sin \Phi -\bigg[R_m \bigg(y \cos(\Delta_f+\delta
\phi_{\lambda_f})-x \sin (\Delta_f+\delta \phi_{\lambda_f})
\bigg)-R^{-1}_m \bigg(y \cos(\Delta_f-\delta
\phi_{\lambda_{\overline f}}) -x \sin (\Delta_f-\delta
\phi_{\lambda_{\overline f}})   \bigg)\bigg] \nonumber\\&\times&
\cos \Phi \bigg\}
\end{eqnarray}
In the models where $r_f$ and $r_{\overline f}$ are so small,
$\delta\phi_{\lambda_f}\simeq \delta \phi_{\lambda_{\overline
f}}\simeq0 $ and hence $A_\Gamma^f$ reduces to
\begin{eqnarray} A_\Gamma^f|_{r_f=r_{\overline f}=0} &=&
\frac{R_f}{2} (R_m+R^{-1}_m) \big(x \cos \Delta_f + y \sin\Delta_f
\big) \sin \Phi - \frac{R_f}{2}
(R_m-R^{-1}_m)\big(y \cos\Delta_f- x \sin \Delta_f\big) \cos \Phi\nonumber\\
\label{apr}\end{eqnarray} This expression is in agreement with the
result in first line of Eq.(23) in Ref.\cite{Grossman:2006jg}. At
tree-level, the amplitudes of the decay processes under concern
have no $\rm CP$ violating weak phases and thus
$\phi^T_f=\phi^T_{\overline f}=0$ implying that $\Phi =\phi -
\phi^T_f-\phi^T_{\overline f}=\phi$. Thus, for $\phi=0$ and $R_m =
1$, i.e. no $\rm CP$ violation in $D^0-\overline D^0$ mixing, we
find that
\begin{eqnarray}
 A_\Gamma^f &=& -\frac{R_f}{2} \bigg\{y \bigg(\cos(\Delta_f+\delta
\phi_{\lambda_f}) - \cos(\Delta_f-\delta \phi_{\lambda_{\overline
f}})\bigg) - x \bigg(\sin (\Delta_f+\delta
\phi_{\lambda_f})-\sin(\Delta_f-\delta \phi_{\lambda_{\overline
f}})\bigg)\bigg\}\label{ef4}\nonumber\\
\end{eqnarray}
The quantities  $\delta \phi_{\lambda_{f}}$ and $\delta
\phi_{\lambda_{\overline f}}$ are expected to be small and thus we
can expand $A_\Gamma^f $ in the preceding equation and keep terms
up to linear order in $\delta \phi_{\lambda_{f}}$ and $\delta
\phi_{\lambda_{\overline f}}$. Thus, we obtain
\begin{eqnarray}
 A_\Gamma^f &\simeq& \frac{R_f}{2} \big(\delta
\phi_{\lambda_f}+\delta \phi_{\lambda_{\overline f}}\big) \big(y
\sin\Delta_f + x \cos \Delta_f\big) \nonumber\\&\simeq& - R_f
\big(r_f \cos\delta_f\sin\phi_f+r_{\overline f}
\cos\delta_{\overline f}\sin\phi_{\overline f}\big) \big(y
\sin\Delta_f + x \cos \Delta_f\big) \label{efwf}\end{eqnarray}

In the SM, the contributions to the amplitudes of the CF $D^0\to
K^-\pi^+$ and the DCS  $D^0\to K^+\pi^-$ decays originate from
integrating out the $W^\pm$ boson mediating the tree-level
diagrams. These contributions are proportional to the Fermi
coupling constant, $G_F$, and CKM matrix elements $V_{U D}$ where
$U=u,c,t$ and $D=d,s,b$. For the scenarios in which subleading
contributions to the tree-level amplitudes arise from new physics
with particles heavier than $m_W$ or from SM amplitudes with
suppressed CKM factors, one finds that $R_f\simeq
|\frac{A^T_{\overline f}}{A^T_{f}}|\simeq
\frac{V_{cd}V^*_{us}}{V_{ud}V^*_{cs}}\simeq \mathcal O (10^{-2})$.
In the SM also, the values of $x,y $ can be as high as $x, y =
\mathcal O (10^{-2})$ \cite{Bigi:2000wn,Falk:2001hx,Falk:2004wg}.
On the other hand, $x$ can be close to the experimental limit in
some classes of NP models
\cite{Grossman:2006jg,Nelson:1999fg,Petrov:2003un}. As a
consequence, we deduce from Eq.(\ref{efwf}) that $A_\Gamma^f $ is
at least suppressed by a factor $x\, R_f \simeq \mathcal O
(10^{-3})$. Other suppression factors can originate from $r_f$ and
$r_{\overline f}$. The weak CP violating and the strong CP
conserving phases may also generate suppression factors in
$A_\Gamma^f $. To obtain upper bound on $|A_\Gamma^f |$, it is
sufficient to find upper bounds on $r_f$ and $r_{\overline f}$
assuming no suppressions from the weak and the strong phases in a
treatment similar to the one adopted in section IV in
Ref.\cite{Grossman:2006jg} although the treatment there is for
direct CP asymmetry.

 As discussed in Ref.\cite{Grossman:2006jg}, and in the presence of
new contributions to the QCD penguin and dipole operators, one can
use the QCD factorization as a framework to obtain
order-of-magnitude estimates for the amplitudes. However, for
hadronic D decays, the $1/m_c$ expansion is not expected to work
very well \cite{Grossman:2006jg}. This can be understood as the
mass of the charm quark is of order 1.5 GeV which is not heavy
enough to allow for a sensible heavy quark expansion like the case
of $1/m_b$ expansion in B meson decay. Thus, as proposed in
Ref.\cite{Grossman:2006jg}, see appendix A for details,  one can
ignore $\mathcal O (\alpha_s)$ corrections to the matrix elements,
as they are negligible compared to the overall theoretical
uncertainties and work primarily at leading order in
$\Lambda_{QCD}/m_c$, using naive factorization (NF) for tree and
QCD penguin operators in the effective Hamiltonian governs the
decay process. Thus, in the following we adopt NF in our analysis
to give an estimation of the upper bounds on $r_f$ and
$r_{\overline f}$ for the CF decay mode $D^0 \rightarrow K^- \pi^+
$ in the framework of the SM and classes of NP models.

\section{ the effective
time-integrated CP asymmetry in the SM \label{dkp}}

In the SM, the total amplitudes of $D^0\to K^-\pi^+$ and $D^0\to
K^+\pi^-$ decay processes can be expressed as
\cite{Delepine:2012xw,Delepine:2017oor}
\begin{eqnarray}
A^{SM}_{D^0\to K^-\pi^+} &=&  -i{G_F\over \sqrt{2}}V_{cs}^*V_{ud}
\left[(a_1+\Delta a_1) X^{\pi^+}_{D^0 K^-} + (a_2+\Delta a_2)
X^{D^0}_{K^-\pi^+} \right],\nonumber\\A^{SM}_{D^0\to K^+\pi^-} &=&
-i{G_F\over \sqrt{2}}V_{us}V^*_{cd} \left[(a_1+\Delta a'_1)
X^{K^+}_{D^0 \pi^-} + (a_2+\Delta a'_2) X^{D^0}_{K^+\pi^-}
\right],\label{am0}
\end{eqnarray}
with  $ X^{P_1}_{P_2P_3}$ is given by
\begin{eqnarray}
X^{P_1}_{P_2P_3}= if_{P_1}\Delta_{P_2P_3}^2
F_0^{P_2P_3}(m_{P_1}^2),\  \Delta_{P_2P_3}^2=m_{P_2}^2-m_{P_3}^2
\end{eqnarray}
here $f_{P}$ is the $P$ meson  decay constant and $F_0^{P_2 P_3}$
is the form factor. In Eq.(\ref{am0}) $a_1 = c_1+ c_2/N_C $ and
$a_2=-( c_2+c_1/N_C)$ where $N_C$ is the color number account for
the tree-level contributions to the amplitudes. These coefficients
originate from integrating out the $W^\pm$ boson mediating the
tree-level diagrams. On the other hand, and in the same equation,
$\Delta a_{1,2}$ and $\Delta a'_{1,2}$ express the contributions
to the amplitudes resulting from integrating out the $W^\pm$ boson
mediating the box and di-penguin diagrams. These loop
contributions are essential for generating the weak phase required
for having non-vanishing $A_{\Gamma}^f$ as the tree-level
contributions are real.  Their expressions are given as
\begin{eqnarray}
\Delta a_1 &\simeq& -{G_Fm_W^2\over \sqrt{2}\ \pi^2V_{c s}^*V_{u
d}N } {\cal B}_x- {G_F \alpha_S\over 4\sqrt{2}
 \pi^3V_{c s} ^* V_{u d}}\left[{\kappa\over 2}\left(1-{1\over N^2}\right)\right]
 {\cal P}_g \nonumber \\
\Delta a_2 &\simeq& -{G_Fm_W^2\over \sqrt{2}\ \pi^2V_{c s}^*V_{u
d} } {\cal B}_x - {G_F \alpha_S\over 4\sqrt{2} \pi^3V_{c s} ^*
V_{u d}}{3  m_{d} m_c\over 8 N }\chi^{D^0} {\cal P}_g
\end{eqnarray}
where $\kappa =(m_D^2+m_K^2)/2+3m_\pi^2/4$ and \be \chi^{D^0} =
{m_{D^0}^2\over (m_c+m_u)(m_s-m_d)},\ee The quantities ${\cal B}_x
$ and ${\cal P}_g$ originate from the box and di-penguin diagrams
respectively and their expressions are given as \bea {\cal B}_x
&=& V_{c D}^*V_{u D} V_{U s}^*V_{U d} f(x_U,\
x_D)\nonumber\\
{\cal P}_g &=&\left[  V_{c D}^*V_{u D}  E_0(x_D)\right] \left[
V_{U s}^*V_{U d} E_0(x_U)\right]\label{Bp}\eea with $U=u,\ c,\ t$
and $D=d,\ s,\ b$, $x_q=(m_q/m_W)^2$ and $f_{UD} \equiv
f(x_U,x_D)$ where \cite{inami}
\begin{eqnarray}
f(x,\ y) ={7xy-4\over 4(1-x)(1-y)} +{1\over x-y}\left[ {y^2\log
y\over (1-y)^2}\left(1-2x+{xy\over 4}\right)-  {x^2\log x\over
(1-x)^2}\left(1-2y+{xy\over 4}\right)   \right] \nonumber
\end{eqnarray}
and the Inami function $E_0(x)$ is given as
\begin{eqnarray}
E_0(x) &=& {1\over 12(1-x)^4}\left[
x(1-x)(18-11x-x^2)-2(4-16x+9x^2)\log(x)\right]
\end{eqnarray}
Turning now to $\Delta a'_{1,2}$ we find that their expressions
are given as
\begin{eqnarray}
\Delta a'_1 &\simeq& -{G_Fm_W^2\over \sqrt{2}\
\pi^2V_{cd}^*V_{us}N }\,{\cal B}'_x - {G_F \alpha_S\over 4\sqrt{2}
\pi^3 V_{cd}^*V_{us}}\left[{q^2\over 2}\left(1-{1\over N^2}\right)
 \right] {\cal P}'_g \nonumber \\
\Delta a'_2 &\approx& -{G_Fm_W^2\over \sqrt{2}\
\pi^2V_{cd}^*V_{us} } \,{\cal B}'_x  - {G_F \alpha_S\over
4\sqrt{2} \pi^3 V_{cd}^*V_{us}}{5m_s m_D^2\over 8N m_d} {\cal
P}'_g \label{dela}
\end{eqnarray}
where the quantities ${\cal B}'_x , {\cal P}'_g$ can be obtained
from the expressions of ${\cal B}_x , {\cal P}_g$, given in
Eq.(\ref{Bp}), via the replacement $d \leftrightarrow s $.

Using $a_1= 1.2 \pm 0.1$, $a_2 = - 0.5 \pm 0.1$ $, |
F_0^{K\pi}(m^2_{D^2})|= 0.5$ and $Arg(F_0^{K\pi}(m^2_{D^2}))=
75^\circ$ \cite{ElBennich:2009da}, $F^{D\pi}_0 (m_K^2)=0.6$, $F^{D
K}_0 (m_\pi^2)=0.75$ \cite{Amhis:2016xyh}, $f_D= 212.15 \pm 1.45
MeV$ \cite{Aoki:2016frl} and $f_K= 157.5 (2.4) MeV$
\cite{Blossier:2009bx,Aoki:2016frl}, we find that $|\Delta a_{1,2}
| \lesssim \mathcal{O} (10^{-8})$  and $|\Delta a'_{1,2} |\lesssim
\mathcal{O} (10^{-6})$ leading to $r_f \lesssim \mathcal{O}
(10^{-8})$ and $r_{\overline{f}} \lesssim \mathcal{O} (10^{-6})$.
Clearly,  $A^{f\,SM}_{\Gamma}$ is suppressed at least by a factor
of $ \mathcal{O} (10^{-10})$ resulting from the product
$x\,R_f\,r_{\overline{f}}$ leading to the prediction
$|A^{f\,SM}_{\Gamma}|\lesssim \mathcal{O} (10^{-10})$.

\section{the effective
time-integrated CP asymmetry in NP models} In this section we
consider two particular extensions of the SM based on their
potentials to enhance CP violation due to the presence of new
complex couplings. The first model is based on extending the
scalar sector of the SM to include new Higgs doublet. The other
model is based on extending the gauge symmetry of the SM to
include new gauge group. In both models, the new interactions can
provide new sources for the weak CP violating phases essential for
CP violation as we showed in our earlier studies in
Refs.\cite{Delepine:2012xw,Delepine:2017oor}. Based on the studies
and due to the strong constraints on the parameter space of the
two models, $r_{\overline{f}}$ are expected to be small compared
to $r_f$ and thus in the following we give an estimation of the
upper bound on $r_f$ only.
\subsection{Models with Charged Higgs contributions}\label{Hig}
Two Higgs doublet models (2HDM) are simple extensions of the SM.
In 2HDM, only the scalar sector of the SM is extended to include
extra Higgs doublet \cite{Haber:1978jt,Abbott:1979dt}. In the
literature, 2HDM have been classified, according to their
couplings to quarks and leptons, into: 2HDM type I, II or III (for
a review see ref. \cite{Branco:2011iw}). The 2HDM type III (2HDM
III) is of a particular interest to our study due to the presence
of complex couplings of Higgs to quarks which are relevant for
generating the desired CP violating weak phases. The model has
five physical mass eigenstates; heavy CP-even Higgs ($H_0$ ),
light CP-even Higgs ($h_0$), CP-odd Higgs ($A_0$) and finally the
charged Higgs ($H^{\pm}$). In the model also,  both Higgs doublets
can couple to up-type and down-type quarks implying that the
couplings of the neutral Higgs mass eigenstates can lead to flavor
violation in neutral currents at tree-level.  As a result, flavor
changing neutral current processes  can be used to strongly
restraint these couplings
\cite{Crivellin:2012ye,Crivellin:2013wna}. We turn now to the
charged Higgs couplings to the quarks. The interaction Lagrangian
in this case is given as \cite{Crivellin:2012ye,Crivellin:2013wna}
\begin{equation}
\mathcal{L}^{eff}_{H^\pm} = \bar{u}_f {\Gamma_{u_f d_i
}^{H^\pm\,LR\,\rm{eff} } }P_R d_i
+ \bar{u}_f {\Gamma_{u_f d_i }^{H^\pm\,RL\,\rm{eff} } }P_L d_i\, ,\\
 \label{Higgs-vertex}
\end{equation}
where \bea {\Gamma_{u_f d_i }^{H^\pm\,LR\,\rm{eff} } } &=&
\sum\limits_{j = 1}^3 {\sin\beta\, V_{fj} \left( \frac{m_{d_i
}}{v_d} \delta_{ji}-
  \epsilon^{ d}_{ji}\tan\beta \right), }
\nonumber\\
{\Gamma_{u_f d_i }^{H^ \pm\,RL\,\rm{eff} } } &=& \sum\limits_{j =
1}^3 {\cos\beta\,  \left( \frac{m_{u_f }}{v_u} \delta_{jf}-
  \epsilon^{ u\star}_{jf}\tan\beta \right)V_{ji}}
 \label{Higgsv}
\eea where $v_u$ and $v_d$ denote the vacuum expectations values
of the neutral component of the  Higgs doublets,  tan $\beta =
v_u/v_d$ and $V$ is the CKM matrix. Extensive study of all
possible constraints that can be imposed on the parameters
$\epsilon^{ u,d}_{jf}$ has been carried in
Ref.\cite{Crivellin:2013wna}. We also have studied the constraints
on $\epsilon^{ u,d}_{jf}$ relevant to the process $D^0\to
K^-\pi^+$ in a previous work in Ref.\cite{Delepine:2012xw}. Based
on our study in Ref.\cite{Delepine:2012xw}, the total amplitude,
including Higgs contribution, can be written as
\begin{eqnarray}
A^{SM+H^\pm}_{D^0\to K^-\pi^+} &\simeq&  -i{G_F\over
\sqrt{2}}V_{cs}^*V_{ud} \left[(a_1+\Delta a^{H^\pm}_1)
X^{\pi^+}_{D^0 K^-} + (a_2+\Delta a^{H^\pm}_2) X^{D^0}_{K^-\pi^+}
\right],\nonumber\\\label{am1111}
\end{eqnarray}
Keeping only the dominant contributions to $\Delta
a^{H^\pm}_{1,2}$ we find that \bea \Delta a^{H^\pm}_1 &\simeq&
\frac {\sin 2\beta m_{d }\epsilon^{
u}_{22}\tan\beta\,\chi^{\pi^+} }{ \sqrt{2} \, G_F  m^2_H v_d},\nonumber\\
\Delta a^{H^\pm}_2 &\simeq&   \frac {\sin 2\beta m_{d }\epsilon^{
u}_{22}\tan\beta\,\chi^{D} }{ 2 \sqrt{2} \, G_F m^2_H v_d\,N}\eea
where \be \chi^{\pi^+} = {m_{\pi}^2\over (m_c-m_s)(m_u+m_d)}\ee A
recent analysis has set the bound $ m_{H^\pm}\gtrsim 600$ GeV
independent of $\tan \beta$ in 2HDM II \cite{Arbey:2017gmh}. This
result has been obtained after considering the most recent
constraints from flavour physics and direct charged and neutral
Higgs boson searches at LEP and the LHC. It should be noted that
the obtained bound must be respected also for the charged Higgs
mass in 2HDM III \cite{Crivellin:2012ye}, Thus, for $\tan\beta =
50$ and $m_{H^\pm} = 600$ GeV we find that \bea \Delta
a^{H^\pm}_1&\simeq&   1.1\times 10^{-3} \, Im(\epsilon^{ u}_{22})I \nonumber\\
\Delta a^{H^\pm}_2 &\simeq&   2.3\times 10^{-3}\, Im(\epsilon^{
u}_{22}) I\label{rfH} \eea where we kept only the imaginary parts
required for generating the weak phases and neglected the real
parts of $\Delta a^{H^\pm}_1$ and $\Delta a^{H^\pm}_2$ as they are
much smaller than the SM contributions and they are not relevant
for generating weak phases. The most dominant constraints on
$Im(\epsilon^{ u}_{22})$ arise from the electric dipole moment of
the neutron \cite{Crivellin:2013wna}.  The resultant bound reads $
-0.16 \lesssim \, Im(\epsilon^{ u}_{22})\, \lesssim 0.16$. Thus,
From Eq.(\ref{rfH}), we find that $ |a^{H^\pm}_{1,2}|$ of
$\mathcal{O} (10^{-4})$.  Thus, we obtain $r_f\lesssim \mathcal{O}
(10^{-4}) $ resulting in this model $|A_\Gamma^{f}|\lesssim
\mathcal{O} (10^{-7})$ which still very small compared to the
current experimental value.
\subsection{A new charged gauge boson as Left Right models}
Possible extensions of the SM  include  models based on the gauge
group $SU(2)_L \times SU(2)_R \times U(1)_{B-L}$
\cite{Pati:1973rp,Mohapatra:1974hk,Mohapatra:1974gc,Senjanovic:1975rk,Senjanovic:1978ev,Beall:1981ze,Cocolicchio:1988ac,Langacker:1989xa,Cho:1993zb,Babu:1993hx}.
In these class of models, new complex couplings can arise due to
the interactions of quarks and leptons with the new charged boson.
In turn, this can  affects  CP violation in meson and lepton
sectors. Previous analyses showed that large direct CP violation
can be generated in the Charm and muon sectors if the mixing
between the left and the right gauge bosons is allowed and the
left-right symmetry is not manifest at unification scale
\cite{Chen:2012usa,Lee:2011kn,Delepine:2012xw}. Motivated by this
finding, we study the impact of the new complex couplings in such
particular setup of LRS model on $A_{\Gamma}^f$ of the decay
process $D^0\to K^-\pi^+$. The charged current mixing matrix can
be parameterized as \cite{Herczeg:1985cx,
Langacker:1989xa,Chen:2012usa}
\begin{eqnarray}
\left(\begin{array}{c}
  W^\pm_L \\
  W^\pm_R
\end{array}\right) =
\left(\begin{array}{cc}
  \cos \xi & -\sin \xi \\
{\rm e}^{i\omega}\sin \xi & {\rm e}^{i\omega}\cos \xi
\end{array}\right)
\left(\begin{array}{c}
  W^\pm_1 \\
  W^\pm_2
\end{array}\right)\simeq
 \left(\begin{array}{cc}
  1 & - \xi \\
{\rm e}^{i\omega}\xi & {\rm e}^{i\omega}
\end{array}\right)
\left(\begin{array}{c}
  W^\pm_1 \\
  W^\pm_2
\end{array}\right)
\end{eqnarray}
here $\xi$ is a mixing angle,  $W^\pm_1$ and $W^\pm_2$ denote the
mass eigenstates and $\omega$ is a weak CP violating phase. This
mixing results in interactions between charged quarks and charged
$W$ bosons that reads
\begin{eqnarray}
{\cal L} &\simeq & -{1\over \sqrt{2}} \bar U \gamma_\mu
\left(g_LVP_L+g_R\xi \bar V^RP_R\right)DW_1^\dagger- {1\over
\sqrt{2}} \bar U \gamma_\mu \left(-g_L\xi VP_L+g_R\bar
V^RP_R\right)DW_2^\dagger
\end{eqnarray}
where $\bar V^R={\rm e}^{i\omega}V^R$. Upon integrating out $W_1$
in the usual way and neglecting $W_2$ contributions, given its
mass is much higher, we can express the total amplitude of $D^0\to
K^-\pi^+$ as

\begin{eqnarray}
A^{SM+LRS}_{D^0\to K^-\pi^+} &\simeq&  -i{G_F\over
\sqrt{2}}V_{cs}^*V_{ud} \left[(a_1+\Delta a^{LRS}_1)
X^{\pi^+}_{D^0 K^-} + (a_2+\Delta a^{LRS}_2) X^{D^0}_{K^-\pi^+}
\right],\nonumber\\\label{am9}
\end{eqnarray}
with \be \Delta a^{LRS}_1 \simeq  {g_R\over g_L }\xi \left(\bar
V^R_{ud}-\bar V^{R*}_{cs} \right) a_1
,\,\,\,\,\,\,\,\,\,\,\,\,\,\,\,\,\,\,\,\,\,\,\,\, \Delta a^{LRS}_2
\simeq  {2 g_R\over g_L }\xi \left(\bar V^R_{ud}-\bar V^{R*}_{cs}
\right) \chi^{D^0} a_2 \label{del1}\ee

The measurement of the muon decay parameter $\rho$, which governs
the shape of the overall momentum spectrum, performed by the TWIST
collaboration \cite{MacDonald:2008xf,TWIST:2011aa} can set
constraint on the left right mixing angle $\xi$. The $\rho$
parameter can be linked to $\xi$ via \cite{MacDonald:2008xf}:

\be \rho \simeq \frac{3}{4}\bigg[1- 2\, \big(\frac{g_R}{g_L}
\xi\big)^2\bigg]\ee

Upon defining $\zeta = \frac{g_R}{g_L} \xi $ and using the TWIST
value, from their latest global fit given in Table VII in
Ref.\cite{TWIST:2011aa}, $\rho = 0.74960 \pm 0.00019$  we obtain

\be   3.7\times 10^{-3} \lesssim \zeta \lesssim 2.3\times
10^{-2}\label{zeta} \ee

which represents the allowed $2 \sigma$ range of the mixing
parameter $\zeta$. We move now to discuss the allowed values of $
\bar V^R_{ud}, \bar V^R_{cs}$. The real parts of these quark
mixing matrix elements will be always suppressed by a factor
$\zeta$ and thus can be neglected compare to the SM contributions.
This will be the case also for the imaginary parts of $ \bar
V^R_{ud}, \bar V^R_{cs}$ where they are also suppressed by the
same factor $\zeta$. However, they provide new source of the
desired weak CP violating phases and thus can not be neglected.

Recently, the authors of Ref. \cite{Alioli:2017ces} have
investigated the possible bounds that can be imposed on the
complex couplings of the $W^\pm$ boson to right-handed quarks
using low-energy precision measurements, flavor physics and
collider physics. These bounds can be applied to the couplings in
general left-right symmetric model that allows mixing between the
charged gauge bosons of the $S U(2)_R$ and $S U(2)_L$ as the one
we consider here. The findings of the study in Ref.
\cite{Alioli:2017ces}, imply that the experimental value of
$(\epsilon '/ \epsilon)_K$ and the stringent bounds on the
electric dipole moment of the neutron can allow $ Im(\bar
V^R_{ud})$ to be as large as $9\times 10^{-4}$. Moreover, the
dominant constraint on $\zeta Im(\bar V^R_{c s})$ arise from the
process $K_L\to \pi^0\, e^+\, e^ -$ and can allow $\zeta \,
Im(\bar V^R_{c s})$ to have a maximum value $7\times 10^{-3}$.
Consequently, with the range of $\zeta$ in Eq.(\ref{zeta}),
$Im(\bar V^R_{c s})$ can have a value $\simeq {\mathcal O }(1)$
without violating the imposed constraints from the process $K_L\to
\pi^0\, e^+\, e^ -$. Taking these values into account, we obtain
$|\Delta a^{LRS}_1|\simeq \mathcal{O} (10^{-2}) $ and $|\Delta
a^{LRS}_2|\simeq \mathcal{O} (10^{-1})$. As a consequence, we find
that $r_f \lesssim \mathcal{O} (10^{-2})$ and hence, in this class
of NP models $|A_{\Gamma}^f|\lesssim\mathcal{O} (10^{-5})$. The
result is one order of magnitude smaller than the experimentally
measured value by LHCb collaborators.

\section{Conclusion \label{CONC}}
In this work we have studied the CP asymmetry in the
time-integrated effective widths, $A_{\Gamma}^f$,  for the Cabibbo
Favored $D^0 \rightarrow K^- \pi^+$  decay process within
different models. This asymmetry is very sensitive to both the
scale and the weak phases of the amplitudes generated from the
radiative corrections to the SM tree-level amplitude or from the
New Physics contributions. In the SM, due to the suppression of
the radiative corrections to the tree-level amplitude, we have
shown that $|A_{\Gamma}^f| \lesssim\mathcal{O} (10^{-10})$ which
is very suppressed compared to the recently measured value
$(A^f_{\Gamma})^{Exp.}= (1.6 \pm 1.0)\times 10^{-4}$. It should be
noted that, this experimental result shows only a $1.6\, \sigma$
deviation from zero and the mode served only as a control channel
in Ref.\cite{Aaij:2017idz}, where its consistence with zero has
been used as a justification of the main result of that paper.

 Regarding the prediction in the framework of 2HDM III, we have
found that $|A_\Gamma^{f\,SM+H^\pm}|\lesssim \mathcal{O}
(10^{-7})$ which is several orders of magnitude smaller than the
current experimental value. Finally, in a general left-right
symmetric models, allowing the mixing between the left and the
right gauge bosons and adopting the scenario that the left-right
symmetry is not manifest at unification scale can lead to a value
$|A_{\Gamma}^f|\lesssim\mathcal{O} (10^{-5})$ which is one order
of magnitude smaller than the experimentally measured value by
LHCb collaborators.

\acknowledgments Part of the work has been done during C.A.
Ramirez's visits at Guanajuato University. C.A.R. wants to thank
the Physics Department of the Guanajuato University for their
hospitality.  D. D. is  grateful to Conacyt (M\'exico) S.N.I. and
Conacyt project (CB-156618), DAIP project (Guanajuato University)
and PIFI (Secretaria de Educacion Publica, M\'exico) for financial
support.
 G.F. is supported by  the research grant
NTU-ERP-102R7701 of National Taiwan University.

\end{document}